\documentclass[a4paper,english]{article}
\usepackage[T1]{fontenc}
\usepackage[latin9]{inputenc}
\usepackage{amsmath}
\usepackage{graphicx}
\usepackage{amssymb}
\usepackage{nomencl}
\providecommand{\printnomenclature}{\printglossary}
\providecommand{\makenomenclature}{\makeglossary}
\makenomenclature
\makeatletter
\date{}
\usepackage{babel}

\makeatother

\makeatother

\usepackage{babel}

\begin{document}

\title{Ion Beam Shepherd for Contactless Space Debris Removal}

%\begin{doublespace}

\author{{\normalsize Claudio Bombardelli}%
\thanks{Research Fellow, ETSI Aeronauticos, Plaza Cardenal Cisneros 3%
} {\normalsize and Jesus Peláez}%
\thanks{Professor, ETSI Aeronauticos, Plaza Cardenal Cisneros 3, AIAA Member%
} \\
 \textit{\normalsize Technical University of Madrid (UPM), Madrid,
E-28040,}{\normalsize {} Spain}\\
 \\
 {\normalsize {} }}
%\end{doublespace}

\maketitle
\begin{center}
(Accepted for publication. Journal of Guidance, Control and Dynamics.
December 2010. Submitted August 2010)
\par\end{center}

\printnomenclature{}

%\begin{doublespace}

\section{Introduction}
%\end{doublespace}

%\begin{doublespace}
$\qquad$The steadily increase of the space debris population is threatening
the future of space utilization for both commercial and scientific
purposes. Since the Sputnik-1 launch in 1957 thousands of satellites
have been delivered to orbit with a current launch rate of about 60
new satellites per year. A considerable fraction of the launched mass,
almost 6000 tons at the time of writing, has remained in orbit producing
more than 15000 trackable objects. In the current situation this number
is growing not only because of newly launched satellites but also
due to on-orbit explosions and accidental collisions among resident
space objects. According to a study by Liou and Johnson \cite{liou2009sensitivity},
even assuming no new satellites were launched, the increase rate of
trackable objects generated by accidental collisions would exceed
the decrease rate due to atmospheric drag decay starting from about
the year 2055. This trend is mostly due to large and massive objects
placed in crowded orbits, that is, at altitudes between 800 and 1000
km and near-polar inclination.
%\end{doublespace}

It is widely agreed that, in order to reduce this threat, not only
newly launched spacecraft and upper stages will need to be deorbited
but also a fraction of the existing ones, calling for active debris
removal operations (ADR). If these operations do not start soon, a
{}``snowball effects'' can take place in which collision-generated
objects will generate new collisions leading to an escalation of the
number of debris in orbit \cite{liou2009sensitivity}. 

The work by Liou and Johnson \cite{liou2009sensitivity} is significant
not only because it analyses the beneficial effects of a planned debris
removal campaign but also because it suggests \textit{what debris
should be targeted} \textit{first}. The preference is put on objects
that are more likely to experience collision and to leave a large
amount of potential debris mass in orbit: the conclusion is that active
debris removal, in order to be effective, should deal with large space
debris in crowded orbits up to about 1600 km altitude. By looking
at the current US-SSN catalogue one finds that there are more than
1000 objects with mass larger than 1 ton in the LEO environment (i.e.
having perigee larger than 2000 km) with a total mass of more than
one third of the total catalogued mass in Earth orbit. The great majority
of these objects are in quasi-circular highly-inclined orbits. Whatever
active removal strategy is chosen, it will clearly need to be able
to deorbit an average 2-ton space object in a reasonable amount of
time and with a reasonable cost in terms of hardware and fuel. This
is especially true if one considers debris removal campaigns in which
a few large objects are removed every year and continuously for a
few decades\cite{liou2009sensitivity}.

Several active debris removal concepts have been proposed ranging
from laser systems (\cite{bondarenko1997prospects}, \cite{phipps1997orion})
to electrodynamic tethers (\cite{cb1},\cite{manu_thesis},\cite{takeichi2006practical}).
Solar sails, which are known to be impractical in LEO, have also been
proposed for reorbiting dead satellite in GEO (\cite{todddebris}).

Once a quick and effective removal method has been devised there still
remains an important technological challenge to overcome: the transmission
of momentum from the removal system to the space debris in order to
carry out the deorbiting (or reorbiting) maneuver. The most obvious
way to do that is to dock the removal system with the target before
the deorbiting starts. This operation can, however, be technologically
complex and very risky. Space debris are non-cooperative objects generally
characterized by a problematic attitude motion (tumbling motion, flat-spin
rotation, large amplitude oscillations etc.) which are not easy to
dock. Another possible solution is to perform a capture operation
with some kind of appendage (e.g. a net or an harpoon) released from
the spacecraft. In this case the major difficulty is perhaps connected
with the deployment and targeting of the capturing device, which,
in addition, would be difficult to reuse for multiple removal operations. 

Debris removal concepts based on pulsed-laser ablation systems do
offer a key advantage in this regard as they can be operated far from
the orbiting target, possibly even from the ground. Unfortunately
though, the small impulse obtained from material ablation cannot be
effective against targets of size exceeding about 20 centimeters \cite{phipps1997orion}.

\begin{figure}[!t]
\centerline{\includegraphics[clip,width=8cm]{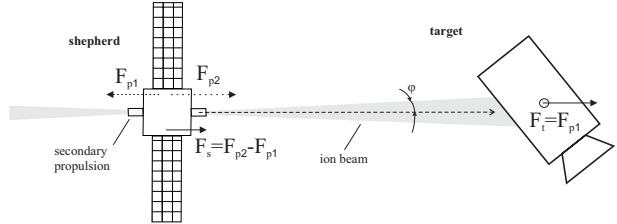}}

\caption{\label{fig:fig1}Schematic of ion beam shepherd satellite deorbiting
a space debris }

\end{figure}

Recently, our team has begun the study of a new space propulsion concept
\cite{bombardelli_patent} in which a highly collimated, high-velocity
ion beam is produced on board a ion beam shepherd spacecraft (IBS)
flying in proximity of a target and directed against the target to
modify its orbit and/or attitude with no need for docking. The momentum
transmitted by the ion beam (ions have been accelerated up to 30 km/s
and more on board spacecraft in past missions) is orders of magnitude
higher than the one obtained, for equal power cost, using material
ablation. Figure \ref{fig:fig1} describes the idea in one of its
most simple implementations. Note that the idea of accelerating a
spacecraft with a flux of incident ions was also recently explored
by Brown et al. \cite{brown2007lunar} who propose a lunar-based ion-beam
generator to remotely propel spacecraft in the Earth-Moon system.
Note also that an independent proposal of using a similar system to
reorbit GEO debris has been put forward by JAXA\cite{Kitamura_debris}. 

%\begin{doublespace}
%

%\end{doublespace}

Potentially, the IBS concept can be used for contactless maneuvering
of space debris irrespectively of their attitude motion. This article
will assess the feasibility of the concept, its expected performance
and its main technological challenges. First the main physics of the
ion-beam momentum propagation are addressed taking into account the
technological level of state of the art ion thrusters. Next, the deorbiting
capability for an optimized system applied to orbital debris in circular
orbit in LEO is evaluated. Additional issues to be addressed in future
studies are outlined and conclusions are drawn.

%\begin{doublespace}

\section{Ion Beam Shepherd Satellite (IBS)}
%\end{doublespace}

The Ion Beam Shepherd concept (IBS) is schematized in Fig.\ref{fig:fig1}.
The shepherd satellite is equipped with a primary propulsion system
that emits a beam of accelerated quasi-neutral%
\footnote{As it is always the case in electric propulsion technology the plasma
leaving the propulsion system is neutralized in order to avoid a net
charge to accumulate on the spacecraft %
} plasma and points it towards a target object in order to apply a
force $\boldsymbol{F}_{d}$ on the latter through the momentum carried
by the plasma ions. If one neglects the momentum associated with ion-sputtering
from the target surface and assuming that the beam fully intercepts
the target, $\boldsymbol{F}_{d}$ will be equal and opposite to the
force $\boldsymbol{F}_{p1}$ that the primary propulsion system exerts
on the shepherd satellite:

\begin{equation}
\boldsymbol{F}_{d}=-\boldsymbol{F}_{p1}.\label{eq:action_reaction}\end{equation}

In the real case, secondary ions and neutrals are sputtered back from
the surface increasing, in principle, the net momentum transmitted
to the target. Yet their ejection velocities are generally small compared
to the ones of the impinging ions \cite{SRIM} so that in the end
the effect on the transmitted force is negligible. On the other hand,
a decrease in the total transmitted momentum occurs when part of the
ions miss the target due to ion beam divergence effects and possible
beam pointing errors, which sets a constraint on the maximum distance
between the IBS and the target as it will be discussed later. Note,
finally, that a misalignment between the beam center of pressure and
the target center of mass does not affect the net momentum transmitted
to the target by the colliding ions as long as the latter continue
to fully intercept the target. This is a consequence of the conservation
of linear momentum of the system before and after the collision. On
the other hand, angular velocity variations do occur in this circumstance
and will be dealt with in future studies.

The magnitude of $\boldsymbol{F}_{p1}$ can then be related to the
primary propulsion system efficiency $\eta_{1}$, the power $P_{1}$
and the ion exhaust velocity $c_{1}$ as:

\begin{equation}
F_{p1}=2\eta_{1}\frac{P_{1}}{c_{1}}.\label{eq:Fp1_mag}\end{equation}

The same quantity can also be related to the mass flow rate $\dot{m}_{1}$
of the propulsion system as:

\begin{equation}
F_{p1}=\dot{m_{1}}c_{1}\label{eq:Fp1_mag2}\end{equation}

The shepherd satellite will then need a secondary propulsion system
(4) to produce an equilibrium force $\boldsymbol{F}_{p2}$ needed
to keep the two satellites at constant distance, as well as a radar
or equivalent measurement system (5) to estimate the position of the
target spacecraft at all time.

In the hypothesis that the IBS and the target debris are in circular
orbit, the magnitude of the force $\boldsymbol{F}_{p2}$ can be computed
by setting to zero the second derivative of the distance $\boldsymbol{d}$
joining the two spacecraft according to:

\begin{equation}
\boldsymbol{\ddot{\textrm{d}}}=\frac{F_{p2}-F_{p1}}{m_{IBS}}-\frac{F_{p1}}{m_{d}}=0,\label{eq:d_dotdot}\end{equation}

where $m_{IBS}$, $m_{d}$ are, respectively, the mass of the debris
shepherd and the mass of the space debris. From the previous equation
one obtains:

\begin{equation}
F_{p2}=F_{p1}\left(1+\frac{m_{IBS}}{m_{d}}\right).\label{eq:Fp2}\end{equation}

The maximum distance $d$ at which the debris shepherd can be held
while the beam fully intercepts the target depends on the size $s$
of the latter and on the ion-beam divergence angle $\varphi$ as: 

\begin{equation}
d_{max}\backsimeq\frac{s}{2\tan\varphi}.\label{eq:d_max}\end{equation}

where $s$ can be thought as the diameter of the largest spherical
envelope contained in the space debris volume.

A simple formula to quantify the smallest divergence angle theoretically
achievable by an ion thruster can be derived from \cite{reiser2008theory}:

\begin{equation}
\varphi_{min}\approx\tan^{-1}\left(\frac{v_{\perp}}{v_{\parallel}}\right)=\tan^{-1}\left(\frac{\sqrt{2q_{e}T_{eV}/m_{i}}}{c}\right),\label{eq:phi}\end{equation}

where $v_{\parallel}=c$ is the exhaust (longitudinal) velocity of
the ions after the acceleration process, $v_{\perp}$ is the rms of
the transverse velocity of the (maxwellian) ions before being accelerated,
$m_{i}$ their mass, $T_{eV}$ the thermodynamic temperature (measured
in electronvolt) of the electrons at the neutralizing cathode (typically
1-5 eV) and $q_{e}$ is the electron charge. The formula highlights
the importance of having high ion ejection velocity (hence high specific
impulse) in order to reduce the beam divergence as much as possible.
The real divergence will be higher due to the complex mutual and external
interaction of the ions, including thermal fluctuations in the plasma
source, non-linear forces due to space-charge fields and possible
plasma instabilities \cite{reiser2008theory}, so that in the end
laboratory experiments are required to estimate the real behavior.

High-potential ion thrusters, such as the proposed DS4G \cite{walker2006initial},
are particularly effective in reaching low divergence angles, thanks
to their very high ion exhaust velocity and the use of multiple grids.
Tests performed by the European Space Agency suggest divergence angles
of 2-5 deg \cite{walker2006initial} which, applied to the present
concept, would allow to remotely control a satellite from a distance
of about 6 to 14 times its size. A space debris of 2 m diameter, for
instance could be controlled with best efficiency at a distance of
12-28 m. More conventional two-grid ion thrusters have a somewhat
larger divergence angle (for example, 15 degrees is the suggested
value for the NSTAR ion thruster \cite{gardner97predictions}) hence
requiring the shepherd to fly much closer to the debris if maximum
efficiency is to be reached. Clearly, in order to relax the control
requirements and minimize collision risks one could employ a control
distance somewhat larger than the one provided by eq. (\ref{eq:d_max})
at the price of a small efficiency loss due to the beam only partially
hitting the target.

%\begin{doublespace}

\section{IBS Mass Optimization for Constant Thrust}
%\end{doublespace}

A design optimization of the IBS will now be performed, in which the
optimum value of the propellant exhaust velocity is derived in order
to minimize the total IBS mass for a debris deorbiting or reorbiting
mission. The optimization process is carried out under the following
assumptions:
\begin{enumerate}
\item The mission begins with the IBS coorbiting with the debris on an initial
generic orbit and ends when the two satellites have reached a common
target orbit. 
\item The thrust provided to the space debris, assumed equal and opposite
to the one provided by the primary propulsion system (Eq. (\ref{eq:action_reaction})),
is constant throughout the mission.
\item The primary and secondary propulsion systems employ ion thrusters
with the same efficiency ($\eta_{1}=\eta_{2}=\eta$) and exhaust velocity
($c_{1}=c_{2}=c$). 
\end{enumerate}
In addition, the notation is simplified by setting:

\[
\dot{m}_{1}=\dot{m};\quad P_{1}=P;\quad F_{p1}=F_{p}.\]

Following the above equalities and using Eqs. (\ref{eq:Fp1_mag},\ref{eq:Fp1_mag2})
the propulsion force $F_{p}$ can be written, for later use as:

\begin{equation}
F_{p}=\dot{m}c=2\eta\frac{P}{c},\label{eq:Fp}\end{equation}

from which one deduces that the power $P$ provided to each thruster
is also constant.

The mass flow rate and power associated with the secondary propulsion
system is computed from the equilibrium condition (Eq. (\ref{eq:Fp2}))
as:

\[
\frac{\dot{m}_{2}}{\dot{m}}=\frac{P_{2}}{P}=\frac{F_{p2}}{F_{p1}}=1+\frac{m_{IBS}}{m_{d}}.\]

The hypothesis that $m_{IBS}\ll m_{d}$ is now introduced, which is
usually reasonable for the case of low-thrust large space debris deorbiting/reorbiting
as it will be confirmed later in the article. Following the above
hypothesis one obtains:

\[
\dot{m}_{2}\thickapprox\dot{m};\quad P_{2}\thickapprox P;\quad F_{p2}\thickapprox F_{p}.\]

The total mass of the IBS is made up by the total propellant mass
($m_{fuel}$) spent throughout the mission duration $\Delta t$, the
power system mass ($m_{p}$) and the structural mass ($m_{str}$).
Since $c$ and $F_{p}$ are constant the former can be easily computed,
with the help of Eq\ref{eq:Fp} as: 

\begin{equation}
m_{fuel}=\intop_{\Delta t}2\dot{m}\textrm{dt}=\intop_{\Delta t}\frac{2F_{p}}{c}\textrm{dt}=\frac{2F_{p}}{c}\Delta t.\label{eq:m_fuel}\end{equation}

Similarly, the power system mass can be computed as:

\begin{equation}
m_{p}=2\alpha P=\frac{\alpha F_{p}c}{\eta}.\label{eq:m_power}\end{equation}

where $\alpha$ denotes the inverse of the specific power ($\mathrm{kg/W}$),
sometimes called {}``specific mass'', of the power generation system.

After summing up the three terms and setting to zero the derivative
with respect to $c$ one obtains the optimum exhaust velocity that
minimizes the total IBS mass:

\begin{equation}
c_{opt}=\sqrt{\frac{2\eta\Delta t}{\alpha}},\label{eq:c_opt}\end{equation}

which is the Irving-Stuhlinger%
\footnote{Note that in Stuhlinger book the thruster efficiency is not accounted
for in the formula and that the specific power, rather than the inverse
if the specific power, is employed%
} characteristic velocity\cite{stuhlinger1964ion}. The corresponding
optimum specific impulse is simply $I_{sp}^{opt}=g_{0}c_{opt}$ with
$g_{0}$ indicating the sea level surface gravity of 9.8$\mathrm{m/s^{2}}$.

Finally the optimized total mass of the IBS becomes:

\begin{equation}
m_{IBS}^{opt}=2F_{p}\sqrt{\frac{2\alpha\Delta t}{\eta}}+m_{str},\label{eq:m_IBS_opt}\end{equation}

while the spent propellant mass yields:

\begin{equation}
m_{fuel}^{opt}=F_{p}\sqrt{\frac{2\alpha\Delta t}{\eta}}.\label{eq:m_fuel_opt}\end{equation}

%\begin{doublespace}

\section{Deorbit Performance }
%\end{doublespace}

A preliminary assessment of the IBS deorbit performance can be done
analytically given the following assumption:
\begin{enumerate}
\item The target debris is in a circular orbit
\item The applied deorbit force is constant, fixed by the mission designer,
and always directed along the tangent to the orbit
\item During the spiral-transfer the orbit evolves in a quasi-circular manner
\end{enumerate}
The assumption are reasonable given the fact that the great majority
of space debris are in almost circular orbits and that the thrust
magnitude achievable with high-performance ion thrusters, typically
less than 200 $\mathrm{mN}$, will produce a negligible variation
of eccentricity when large debris pieces ($m_{d}\gtrsim1000$ kg)
are considered.

For a generic orbit, the time evolution of the orbit semimajor axis
$a$ under the tangential perturbing force $F_{p}$ obeys the Gauss
equation:

\begin{equation}
\frac{da}{dt}=\pm\frac{2a^{2}v}{\mu}\frac{F_{p}}{m_{d}},\label{eq:gauss}\end{equation}

where $\mu$ is the earth gravitational constant, $v$ the space debris
velocity and the sign - (+) indicates deorbit (reorbit). Under the
hypothesis that the orbit evolves while remaining almost circular
($v=\sqrt{\mu/a}$ ), Eq. (\ref{eq:gauss}) can be replaced by:

\begin{equation}
\frac{da}{dt}=\frac{2a^{3/2}}{\mu^{1/2}}\frac{F_{p}}{m_{d}}.\label{eq:Eq_gauss_circ}\end{equation}

Since $F_{p}$ is constant, Eq. \ref{eq:Eq_gauss_circ} can be integrated
to provide the orbit radius evolution in time, which for the case
of drag and thrust, respectively, yields:

\begin{equation}
a_{deorb}=\frac{\mu R}{\left(\dfrac{F_{p}}{m_{d}}t\sqrt{R}+\sqrt{\mu}\right)^{2}}\label{eq:deorb}\end{equation}

\begin{equation}
a_{reorb}=\frac{\mu r}{\left(\dfrac{F_{p}}{m_{d}}t\sqrt{R}-\sqrt{\mu}\right)^{2}}\label{eq:reorb}\end{equation}

where $R$ and $r\mbox{ }$ indicate, respectively, the radius at
the beginning of the deorbit and reorbit maneuver.

The time duration of the maneuver is obtained by solving Eq. (\ref{eq:deorb})
and (\ref{eq:reorb}) for $t$ after setting $a_{deorb}=r$ and $a_{reorb}=R$.
In both cases the time span obeys: 

\begin{equation}
\Delta t=m_{d}\frac{\sqrt{\mu}}{F_{p}}\times\frac{\sqrt{R}-\sqrt{r}}{\sqrt{rR}}.\label{eq:deorbit_time}\end{equation}

After substituting Eq. (\ref{eq:deorbit_time}) into Eq. (\ref{eq:total_power})
one finally obtains the total mass of the optimized IBS system for
maneuvering a space debris of mass $m_{d}$ between circular orbits
of radii $r$ and $R$ with constant tangential low thrust of magnitude
$F_{p}$ : 

\begin{equation}
m_{IBS}^{opt}(m_{d},r,R,F_{p})=2\left(\frac{\mu}{Rr}\right)^{1/4}\sqrt{\frac{2\alpha m_{d}F_{p}}{\eta}\left(\sqrt{R}-\sqrt{r}\right)}+m_{str}\label{eq:m_IBS_opt_orb}\end{equation}

The propellant mass spent throughout the deorbiting maneuver is computed
by substituting Eq.(\ref{eq:deorbit_time}) into Eq.(\ref{eq:m_fuel_opt})
to yield:

\begin{equation}
m_{fuel}^{opt}(m_{d},r,R,F_{p})=\left(\frac{\mu}{Rr}\right)^{1/4}\sqrt{\frac{2\alpha m_{d}F_{p}}{\eta}\left(\sqrt{R}-\sqrt{r}\right)}.\label{eq:m_fuel_opt_orb}\end{equation}

Finally, the total power needed by the optimized system can be derived
from Eq. (\ref{eq:Fp}) and taking into account Eqs. (\ref{eq:c_opt},\ref{eq:deorbit_time}):

\begin{equation}
P^{opt}(m_{d},r,R,F_{p})=\left(\frac{\mu}{Rr}\right)^{1/4}\sqrt{\frac{2m_{d}F_{p}}{\eta\alpha}\left(\sqrt{R}-\sqrt{r}\right)}.\label{eq:total_power}\end{equation}

\begin{figure}[!t]
\centerline{\includegraphics[clip,width=8cm]{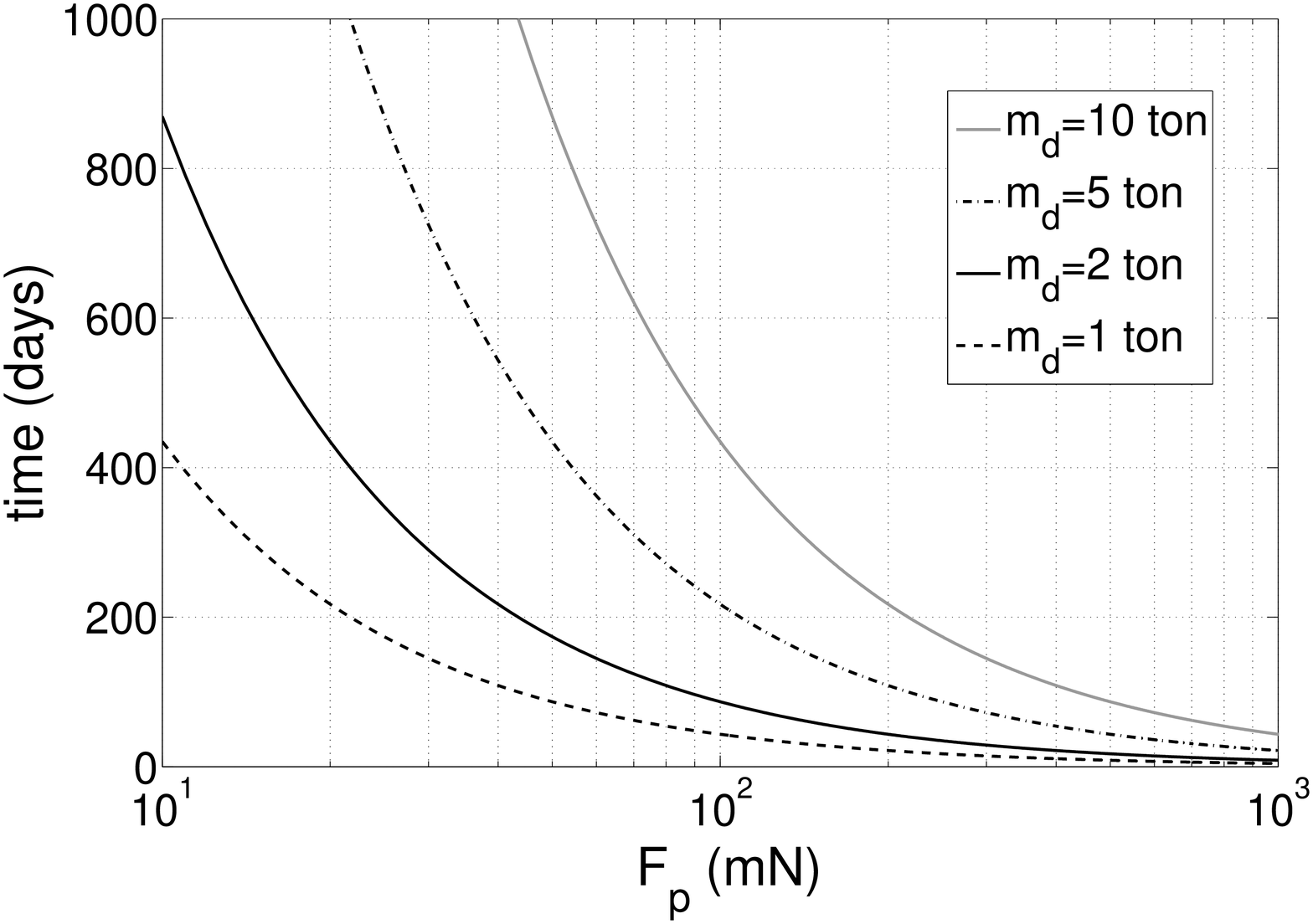}}

\caption{\label{fig:fig2}Time required for transferring space debris of different
masses $m_{d}$ from a 1000-km- to a 300-km-altitude circular orbit
with an IBS providing constant tangential thrust $F_{p}$. }

\end{figure}

\begin{figure}[!t]
\centerline{\includegraphics[clip,width=8cm]{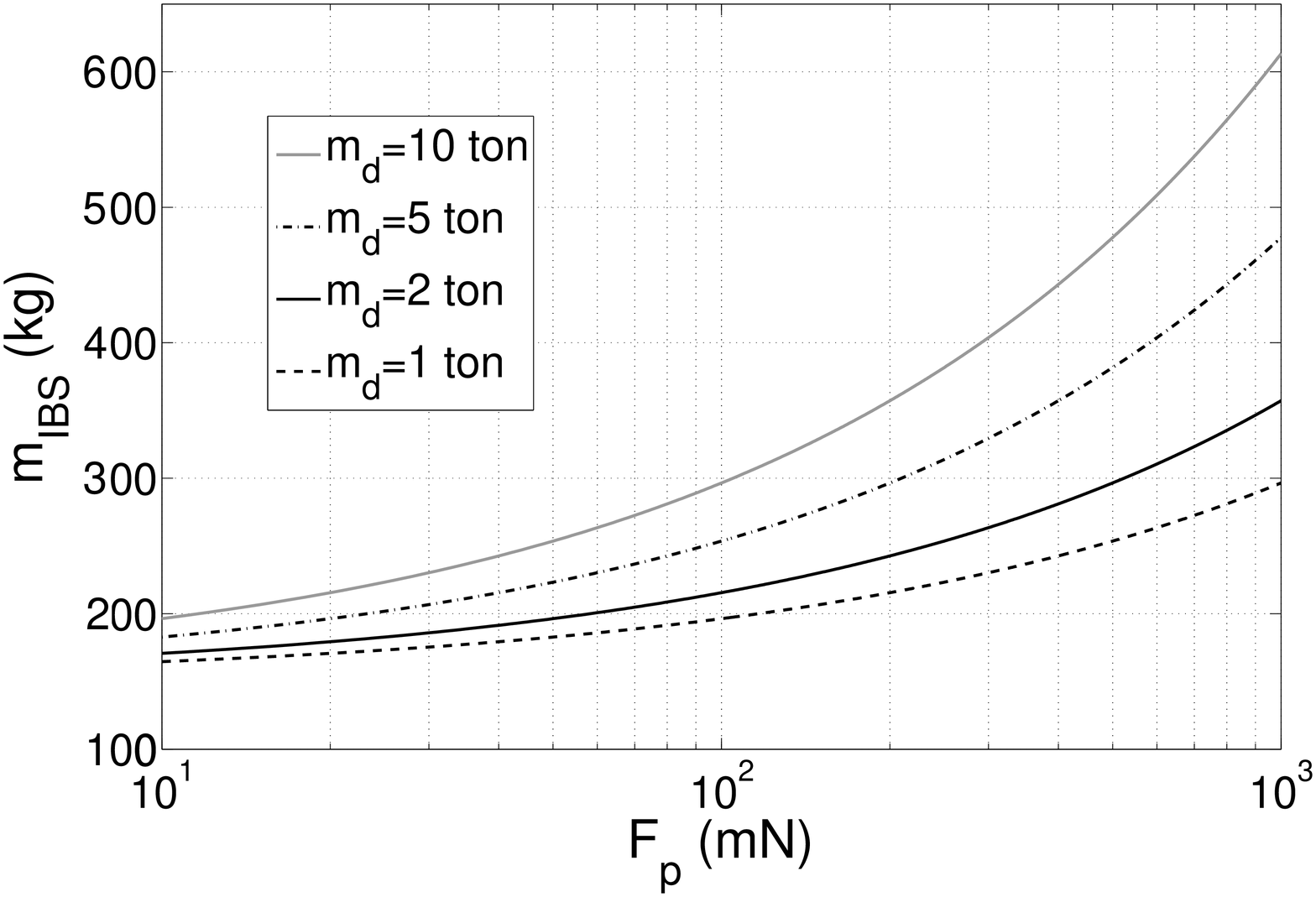}}\centerline{\includegraphics[clip,width=8cm]{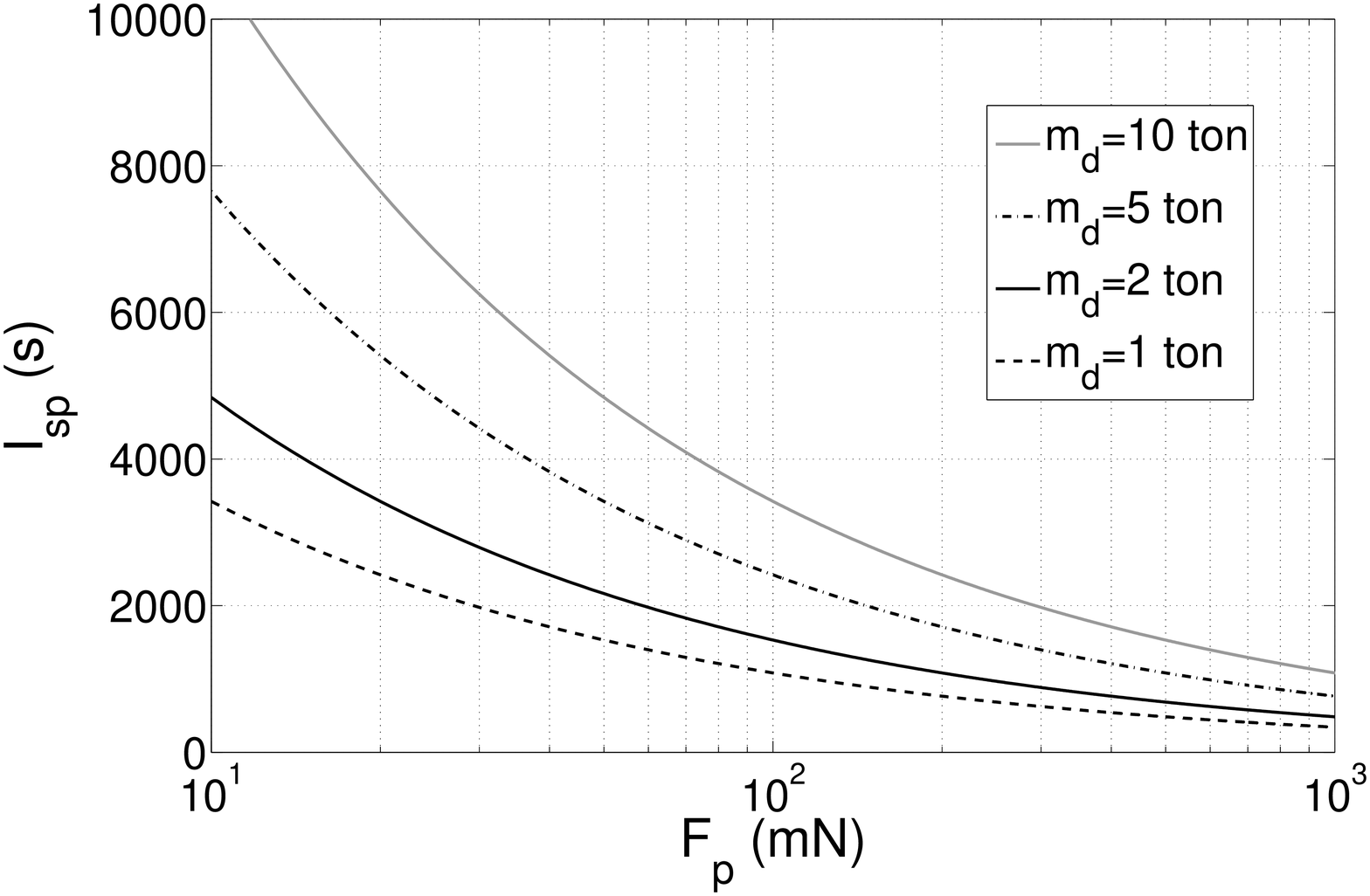}}\centerline{\includegraphics[clip,width=8cm]{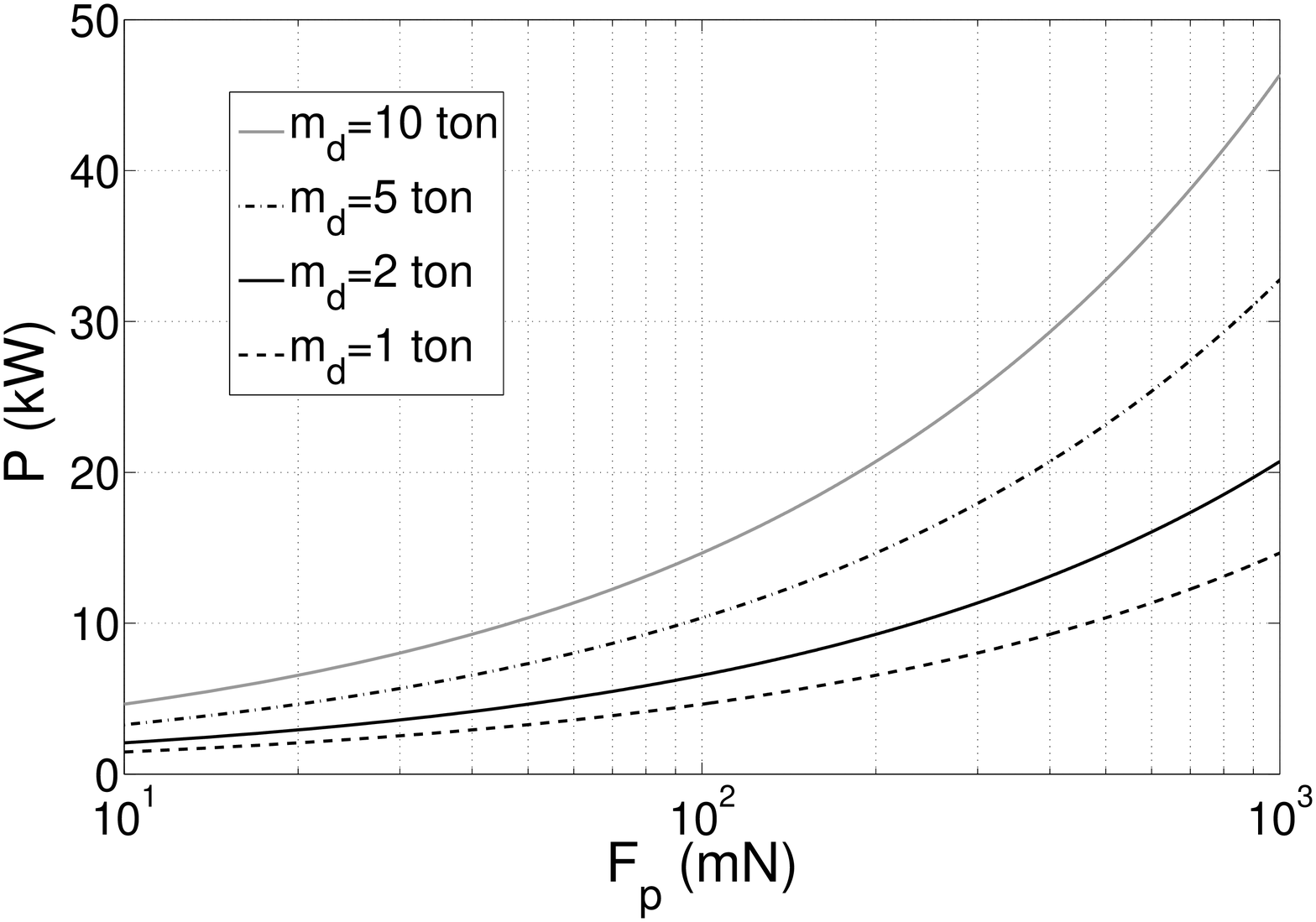}}

\caption{\label{fig:fig3}Total optimized IBS mass (top), specific impulse
(middle) and power (bottom) required for transferring space debris
of different mass $m_{d}$ from a 1000-km- to a 300-km-altitude circular
orbit with an IBS providing constant tangential thrust $F_{p}$. Two
equal ion engines with $\eta=70\%$ and optimum ejection veleocity
(Eq.(\ref{eq:c_opt})) are employed as primary and secondary propulsion
system. The common power plant feeding the two ion engines has specific
mass $\alpha=5\mathrm{kg/kW}$. A total structure mass of 150 kg is
assumed. The ion beam is assuemd to fully intercept the target throughout
the maneuver. Note that both the power system mass and the propellant
mass spent can be obtained by subtracting the structure mass to the
total spacecraft mass and dividing by two.}

\end{figure}

Figures \ref{fig:fig2} and \ref{fig:fig3} plot the deorbit time
(Eq.(\ref{eq:deorbit_time})) and the optimized IBS mass (Eq. (\ref{eq:m_IBS_opt_orb}))
required to transfer space debris of different sizes from a circular
orbit of 1000 km altitude (a high-density debris orbit) to a lower
300-km-altitude circular orbit (below the International Space Station).
It can be seen that, for instance, a 100 mN thrust ion thrust with
70\% thrust efficiency and employing a power plant with $\alpha=5$
kg/kW is capable of deorbiting a 5-ton debris in less than one year
with less than 300 kg total spacecraft mass by employing ion thrusters
with an optimized specific impulse $I_{sp}\sim2500$s. Note, however,
that in order to reduce the beam divergence, allowing a higher control
distance between the shepherd and the target debris, a higher specific
impulse compared with the mass-optimum value may be desirable resulting
in a small increase in total system mass. This kind of design and
optimization trade-off, which requires experimental data on thruster
plume divergence for different values of the specific impulse, is
beyond the scope of the present note. 

%\begin{doublespace}
%

%\end{doublespace}

%\begin{doublespace}

\section{Additional Issues }
%\end{doublespace}

While the present study has shown that the IBS concept is a promising
solution for active debris removal further investigation is needed
to address the following issues:

- Ion beam momentum transmission under non-nominal conditions: Analytical
and numerical models will be needed to compute the force transmitted
to a target once the constraint \ref{eq:phi} is not satisfied.

- Proximity formation flying control: The relative dynamics and control
of an ion-beam-propelled target space debris need to be investigated.

- Attitude dynamics of the target: As a consequence of the misalignment
of the ion-beam center of pressure and target center of mass a net
torque originates affecting the target attitude dynamics. While the
total linear momentum transmitted to the target, hence the deorbiting
efficiency, is not influenced by this effect an excessive spin-up
of the target could pose operational risks (e..g centrifugal fragmentation)
so that a proper control strategy is likely necessary.

- The flux of secondary ions backscattered from the target surface
needs to be estimated in order to address possible risks of contamination
of sensitive parts (e.g. solar panels, electronics) of the shepherd
satellite. 

%\begin{doublespace}

\section{Conclusions}
%\end{doublespace}

A new concept for active removal of space debris has been presented,
in which a space debris shepherd uses the momentum transmitted by
a low-divergence accelerated ion beam in order to achieve contactless
debris removal. A preliminary analysis of the concept has been conducted
highlighting the key aspects of the system design and its performance.
Ion thrusters with low beam divergence (< 15 deg), available from
current space hardware, are key to allow contactless maneuvering at
safe distance from the debris. A design optimization has been conducted
in order to minimize the required total mass of the debris shepherd
showing that, in the hypothesis that the former is much smaller than
the debris mass, optimum specific impulse corresponds to thrust exhaust
velocity equal to the Irving-Stuhlinger characteristic velocity. The
deorbiting performance for large size (>1 ton) debris has been evaluated
analytically in the hypothesis of quasi-circular orbit evolution.
As a numerical example, a very large (5 tons) space debris can be
deorbited in about 7 months with a total IBS mass of less than 300
kg assuming, as a very preliminary value, a structural mass of 150
kg.

Although the concept implementation appears to be feasible with state
of the art space hardware further analysis will be required to investigate
the physical interaction between an orbiting body and an ion beam
including sputtering phenomena and possible plasma backflow from the
debris surface towards the IBS spacecraft. The control of the relative
distance between the IBS and the debris flying in proximity for a
large time span and the attitude dynamics of the debris will also
need to be addressed.

\section{Acknowledgments}

The work for this paper was supported by the {}``ARIADNA Call for
Ideas on Active Debris Removal'', established by the Advanced Concepts
Team of the European Space Agency and by the reserach project \textquotedblleft{}Propagation
of Orbits, Advanced Orbital Dynamics and Use of Space Tethers\textquotedblright{}
supported by the Dirección General de Investigación (DGI) of the Spanish
Ministry of Education and Science through the contract ESP2007-64068.
We would also like to thank the ESA/ESOC Space Debris Office for kindly
providing statistical data on actual space debris in LEO. 

\medskip{}

%\begin{doublespace}
\bibliographystyle{aiaa}
\bibliography{library}
%\end{doublespace}

\end{document}